\documentclass[conference]{IEEEtran}
\IEEEoverridecommandlockouts
\usepackage{cite}
\usepackage{amsmath,amssymb,amsfonts}
\usepackage{algorithmic}
\usepackage{graphicx}
\usepackage{textcomp}
\usepackage{xcolor}
\def\BibTeX{{\rm B\kern-.05em{\sc i\kern-.025em b}\kern-.08em
    T\kern-.1667em\lower.7ex\hbox{E}\kern-.125emX}}
\begin{document}

\title{HIVA: Holographic Intellectual Voice Assistant\\

}

\author{\IEEEauthorblockN{Ruslan Isaev}
\IEEEauthorblockA{\textit{Department of Computer Science} \\
\textit{Ala-Too International University}\\
Bishkek, Kyrgyzstan \\
ruslan.isaev@alatoo.edu.kg}

\and
\IEEEauthorblockN{Radmir Gumerov}
\IEEEauthorblockA{\textit{Department of Computer Science} \\
\textit{Ala-Too International University}\\
Bishkek, Kyrgyzstan \\
radmir.gumerov@alatoo.edu.kg}

\and
\IEEEauthorblockN{Gulzada Esenalieva}
\IEEEauthorblockA{\textit{Department of Computer Science} \\
\textit{Ala-Too International University}\\
Bishkek, Kyrgyzstan \\
gulzada.esenalieva@alatoo.edu.kg}

\\


\and

 \hspace{4cm}
\IEEEauthorblockN{Remudin Reshid Mekuria}
\hspace{6cm}
\IEEEauthorblockA{\textit{~~~~~~~~~~~~~~~~~~~~~~~~~~~~~~~Department of Computer Science} \\
\hspace{4cm}
\textit{Ala-Too International University}\\
\hspace{4cm}
 Bishkek, Kyrgyzstan \\
\hspace{4cm}
remudin@alatoo.edu.kg}

\and

\IEEEauthorblockN{Ermek Doszhanov}
\IEEEauthorblockA{\textit{Department of Computer Science} \\
\textit{Ala-Too International University}\\
Bishkek, Kyrgyzstan \\
ermek.doszhanov@alatoo.edu.kg}
}

\IEEEoverridecommandlockouts
\IEEEpubid{\makebox[\columnwidth]{979-8-3503-3961-1/23/\$31.00~\copyright2023 IEEE\hfill} \hspace{\columnsep}\makebox[\columnwidth]{ }}
\maketitle
\IEEEpubidadjcol

\begin{abstract}
Holographic Intellectual Voice Assistant (HIVA) aims to facilitate human computer interaction using audiovisual effects and 3D avatar. HIVA provides complete information about the university, including requests of various nature: admission, study issues, fees, departments, university structure and history, canteen, human resources, library, student life and events, information about the country and the city, etc. 
There are other ways for receiving the data listed above: the university's official website and other supporting apps, HEI (Higher Education Institution) official social media, directly asking the HEI staff, and other channels. However, HIVA provides the unique experience of "face-to-face" interaction with an animated 3D mascot, helping to get a sense of 'real-life' communication. The system includes many sub-modules and connects a family of applications such as mobile applications, Telegram chatbot, suggestion categorization, and entertainment services. The Voice assistant uses Russian language NLP models and tools, which are pipelined for the best user experience.
\end{abstract}

\begin{IEEEkeywords}
voice assistant, natural language processing, software engineering, visual question answering, text mining
\end{IEEEkeywords}

\section{Introduction}
Voice assistants are digital helpers that are widely used nowadays for different purposes. They use speech recognition, speech synthesis, and natural language processing technologies to imitate real dialogue i.e. interact with users through voice  and respond with human-like speech. The most known examples are Alexa, Siri, and Google Assistant. Moreover, not only Ala-Too International University (AIU) tends to develop its own voice assistant, but recently Prashn web application was released and described in research \cite{Sharma}. Needless to say that in today's world, working with information in the form of machine and human-readable data has become a cornerstone feature of any industry. Fast access to up-to-date and precise information about an organization is one of the key parameters for the success of an institution. Agility in reply and correctness of response helps build better relationships between an organization and its clients. 

\begin{equation} \label{qasearch}
    \operatorname{ans}[{\mathtt{A}}_i,|,{\mathtt{S}}_i] = \frac{\left|\operatorname{L}\cap\operatorname{set}(a)\right|}{\left|\operatorname{L}\right|} + \frac{\left|\operatorname{set}(a)\cap\operatorname{L}\right|}{\left|a\right|}
\end{equation}

Last year's pandemic has affected in-person communications among people as well. Many global project teams propose various ways of dealing with the lack of personal "real-life" interactions and communications. Almost all of them use new technologies, such as programming, working with big data, machine learning and AI, 3D modeling, virtual and augmented reality \cite{Liu}. The idea of the HIVA project lies in the grounds of the need to modernize orientation processes in the ecosystem of AIU.
 
Nowadays, most of the information is received by interacting with a machine, i.e. computers, smartphones, tablets, and other devices. 3D Holography and Artificial Intelligence technologies can be considered the next level of person-to-machine interaction. Imitating in-person communication as a dialog brings a more immersive experience to a communicator \cite{Conner}. We expect HIVA to help anticipate and reshape the future and expand boundaries for incorporating university-made technologies into real business processes.

\subsection{Examples of AI and 3D Holography engagement}

Today, after receiving big progress in the development of AI systems, many technological startups and businesses implement AI models, i.e. machine learning and artificial neural networks. We can see a boost in AI model usage, which has increased enormously over the past few decades. As a result, we can see a lot of AI assistants appeared in the market \cite{Patel}. 

Having two technologies combined opens new opportunities for various applications including business and educational assistants. Here are some examples of AI-enhanced 3D holography applications:

\begin{enumerate}
\item Human-Computer Interaction: 3D holography as an immersive virtual environment for human-computer interaction, where hand gestures allow users to interact with virtual objects in real-time \cite{Caggianese}.
\item Medical Visualization: realistic, interactive visualizations of medical data, such as 3D models of the human anatomy, modeling 3D organs based on individual patient's medical image data is indispensable for simulation, navigation, and education for accurate and safe surgery \cite{Sugimoto}.
\item AR/VR: augmented and virtual reality experiences, where users can interact with virtual objects in a real-world environment \cite{Caggianese}.
\item Marketing and Advertising: unique and engaging advertisements, such as holographic product displays in retail environments \cite{Patel}.
\end{enumerate}

It's worth noting that these are still early applications of 3D holography, and much more research and development is needed to realize its potential fully. We must mention, that almost all the current AI system lacks interactive part, and scope to visualize the output is limited due to the complexity of technology \cite{Patel}.


\subsection{What holographic effect is used in HIVA, and how does it work?}

Holography is a technique for recording and reproducing three-dimensional images. Some of us may be familiar with effect from science fiction movies such as "Star Wars" or "Blade Runner" where video message is reproduced as a 3D image. Modern artists use holographic effects to create audio-visual installations \cite{Lemercier}. The real holographic effect involves using laser beams and other optical devices to record an interference pattern of light on a photographic plate or other medium. The current research state shows us that we are still at the beginning of the way to real-time animated holography, which may be used to see objects with volume and presence effects. Building a real working device based on actual 3D holography is complicated and will not provide the desired effect. In the HIVA project, we focused on building a holographic effect known as Pepper's ghost \cite{Conner}.

Pepper's Ghost is an illusion technique used in theatre, concerts, and haunted attractions. It was originally developed in the 19th century by John H. Pepper and involves using a sheet of glass or clear plastic to reflect an object or person, making it appear as if they are floating in mid-air. The technique works by reflecting a hidden object or person onto the transparent surface while carefully positioning the audience's line of sight to create the illusion of a ghostly presence \cite{Conner}. It is widely used today, often in combination with modern technology such as projection mapping and LED lighting.


\section{Methodology}

The study involved potential applicants seeking information about the AIU. Data Collection: Data was collected through the use of the HIVA. The HIVA provided information on the AIU and answered applicants' questions. Participants were allowed to access information about AIU through traditional methods, such as the website and social media channels, or by contacting the AIU staff directly or using the HIVA. Participants who chose to use the HIVA were provided with a user manual and given time to familiarize themselves with the system before proceeding with the study. Data collected from participants were analyzed using descriptive statistics, including means and percentages. Participants were informed of the purpose of the study, and their consent was obtained before proceeding with the study. All data collected was kept confidential and anonymous.

The combination of 3D holographic pyramid technology and voice assistant in an HEI-used device makes this project unique for Kyrgyzstan and Kyrgyzstan HEIs. The visual effect that the pseudo-hologram provides is an additional channel of interaction with the user \cite{Conner}. This project has the capacity for extension: HIVA can be modified and used to automate issuing certificates and receiving requests. Today the most common information and reference systems provide information in the form of static information stands or interactive screens. Some research prototypes of interactive holographic pyramids exist and are widely used in museums, training centers, etc. However, interaction with a hologram via voice has not yet been implemented in any project in Kyrgyzstan.

According to project objectives we planned to efficiently receive and process requests of various types from stakeholders. Provide consultation services and general information about AIU in multiple languages. Maintain an up-to-date FAQ section about the admission process and student activities for applicants, students, and their parents. Offer a virtual tour of the university.
The primary performance indicator for the project was the successful completion of the outlined objectives using Agile and Scrum principles.




To effectively develop system within Agile software development (ASD) is based on the Agile Manifesto, a group of developers created in 2001. The Agile Manifesto consists of four values, and twelve principles, which serve as the foundation for ASD methods \cite{Abrahamsson}.

ASD is an iterative and flexible approach that values collaboration, responsiveness to change, and building working software as early as possible. Whereas Agile is a general term based on the Agile Manifesto values and principles, Scrum is just one of several Agile frameworks. 

Scrum is a specific ASD framework that provides a structured approach for managing and completing complex projects. It was originally designed for software development but has since been adapted for various projects and industries. Scrum emphasizes adaptability, collaboration, and delivering working software in small, incremental "sprints" \cite{Srivastava}.

We chose Agile in work over HIVA because it provides a flexible framework (Scrum) for planning, executing, and adjusting the development process and helps respond to changing requirements and regularly arising obstacles and errors. 

\begin{figure}
\centering
\includegraphics[width=0.48\textwidth]{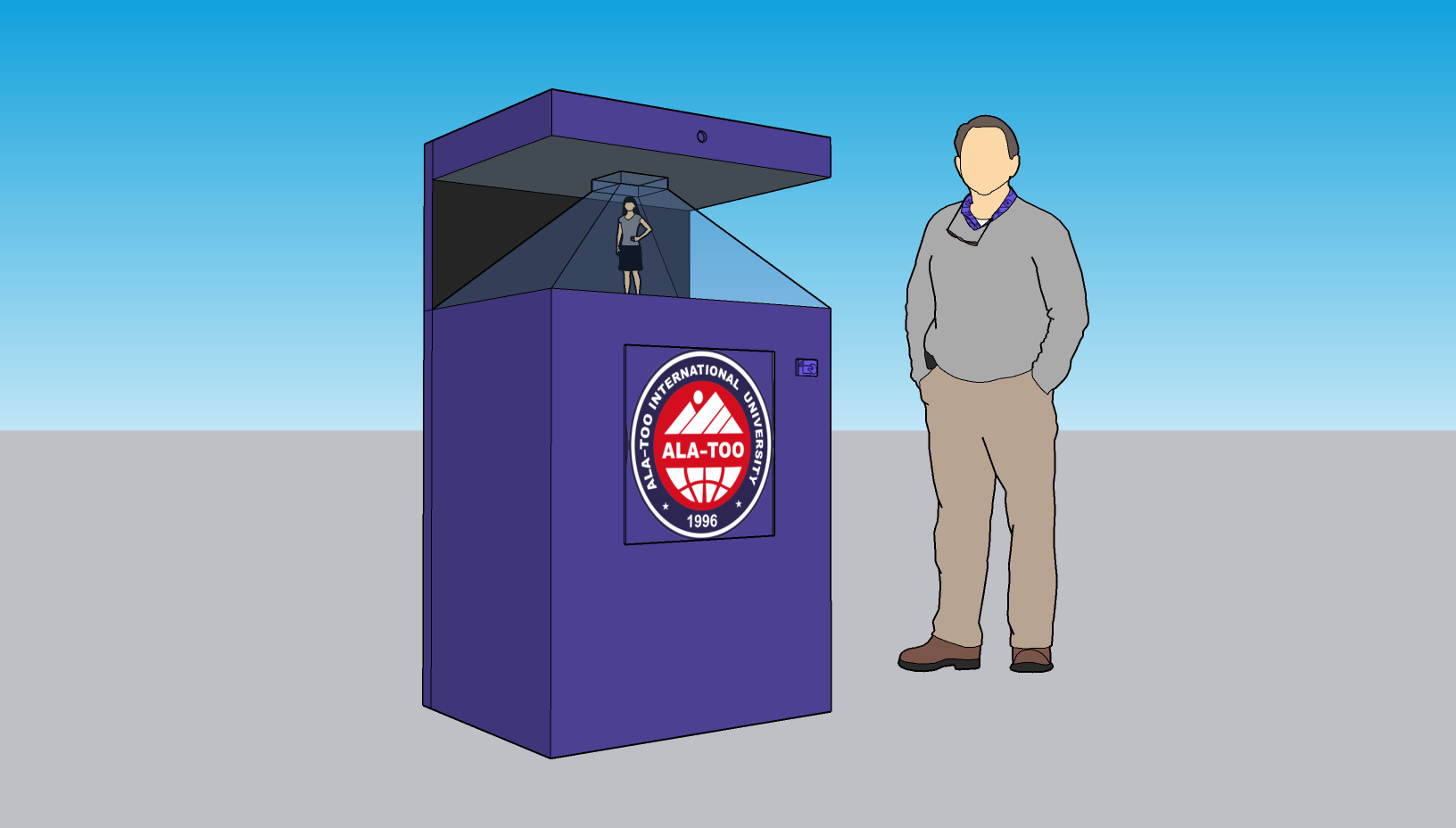}
\includegraphics[width=0.48\textwidth]{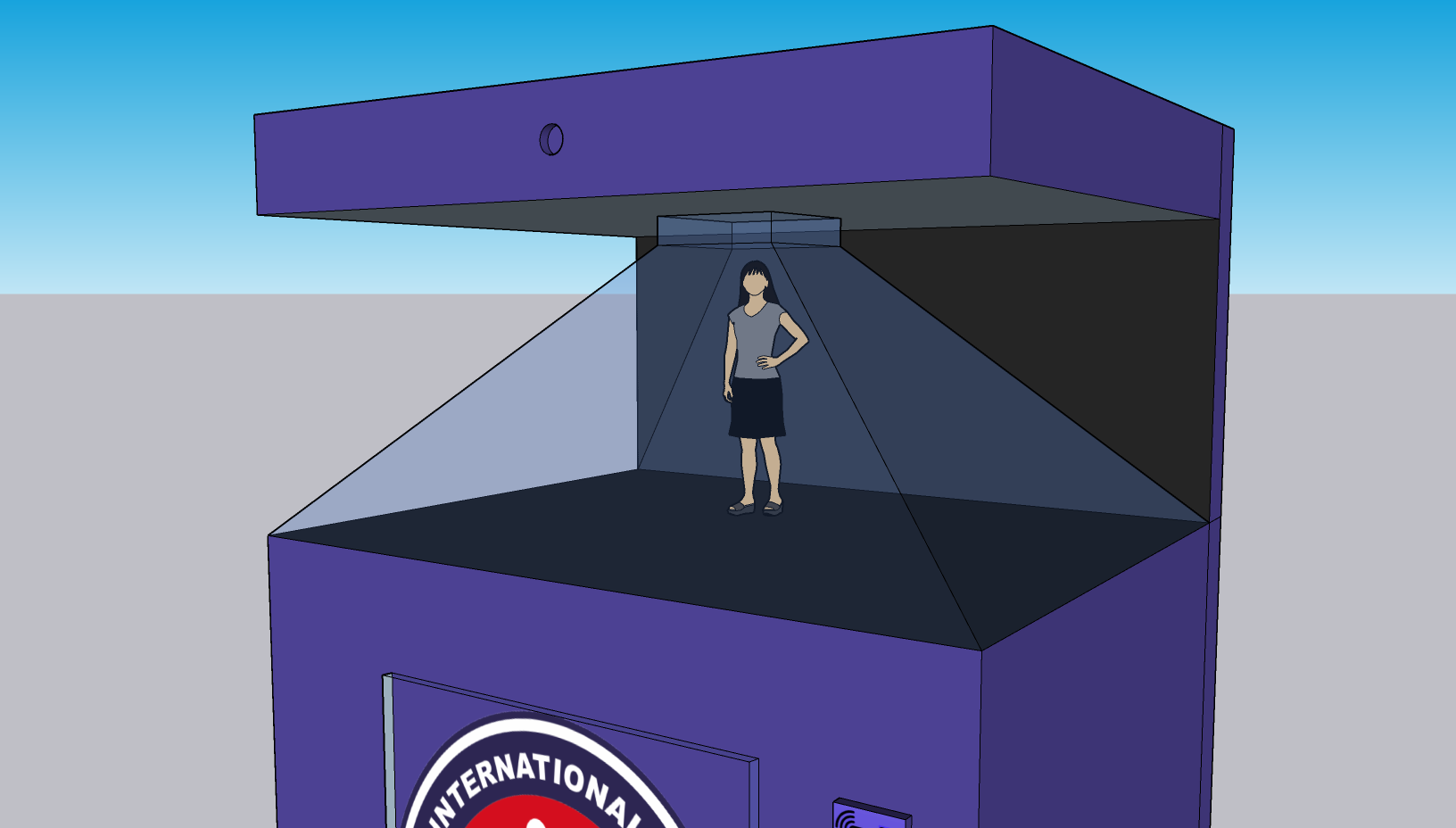}
\caption{Early HIVA prototype design}
\label{fig1:prototype_early}
\end{figure}

The first step (sprint) was dedicated to designing and assembling the device case. The first prototype design is seen in images \ref{fig1:prototype_early}. We used Oculus Quest 2 VR Headset to see the virtual prototype immersed in an environment with a not yet manufactured device to see it from all angles and even "touch" it. During the second step, the project team designed and ordered a holographic pyramid and purchased electronic and computing components. During step three, they embedded a cooling system and an automatic on/off system inside the HIVA. To receive the effect of holography, they have installed and programmed a backlight system. As the first animated avatar for the first version of HIVA, we used the "EVE" model with simple animations (credits to \cite{lucasquincoses}). 

The first HIVA prototype was launched on July 1, 2021. That was the day of the start of admission to AIU. During the tests, project members conducted an anonymous, depersonalized data collection of requests. The collected results were further processed and used as a model for training a dialog system for the next prototype software versions. The question-answer model was manually prepared using a knowledge base from the Frequently Asked Questions section on the official website. 

\begin{figure}
\centering
\includegraphics[width=0.48\textwidth]{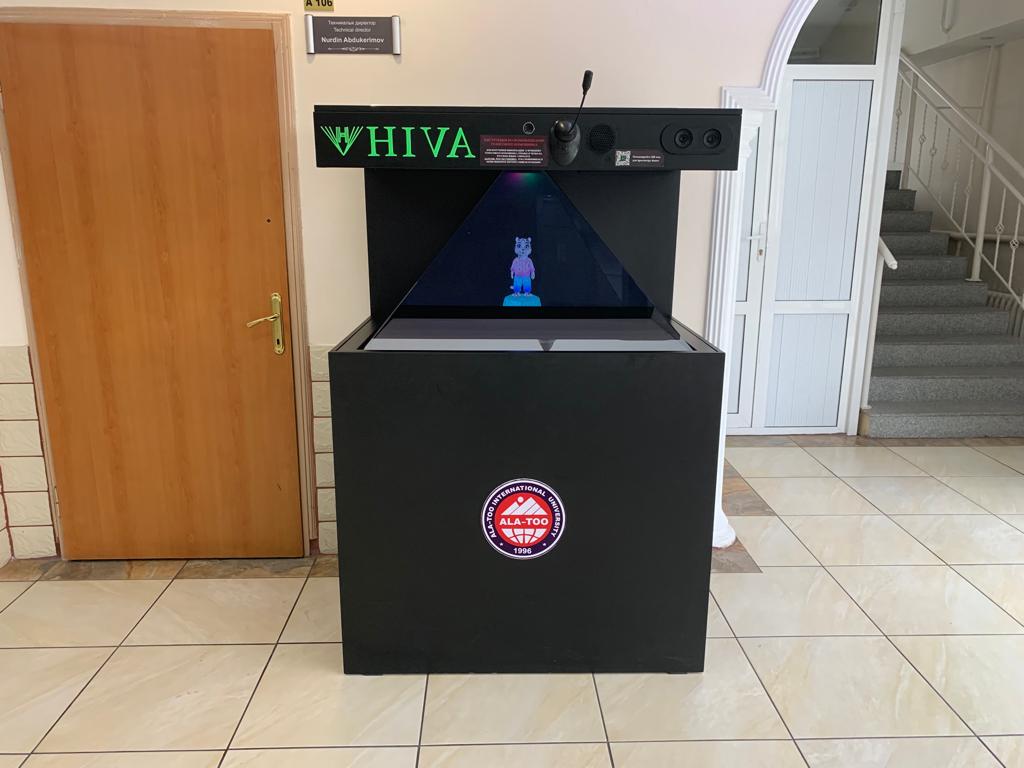}
\includegraphics[width=0.48\textwidth]{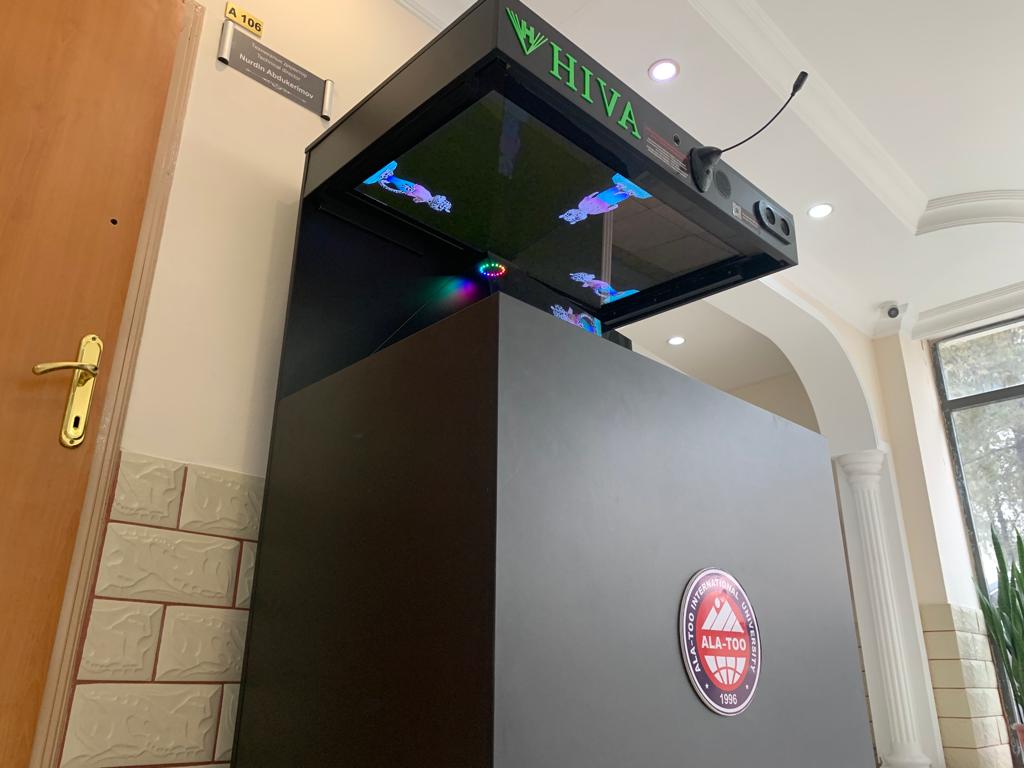}
\caption{HIVA as a working device}
\label{realhiva}
\end{figure}

The work on the project is still in progress, and after releasing the HIVA version, $2.0$ our team already has plans and backlog tasks for HIVA version $3.0$ development. Currently, HIVA works without any Internet access \ref{realhiva}. The new version can recognize and synthesize Russian language speech, search in the database, and provide information about news, weather, and time in Kyrgyzstan. It can play some music as well. You can add this feature to your voice assistant, adding pafy and youtubesearchpython Python libraries as dependencies. Those two provide API to YouTube. Important to know that some dependencies may crash your project since services such as YouTube may change the data structure. The service removed them on the 13th of December 2021, which caused a system failure when a user attempted to execute "music play" using voice command. We fixed this bug by commenting out the dislikes variable from the source code of pafy library. 



\begin{figure}[h!]
\centering
\includegraphics[width=0.48\textwidth]{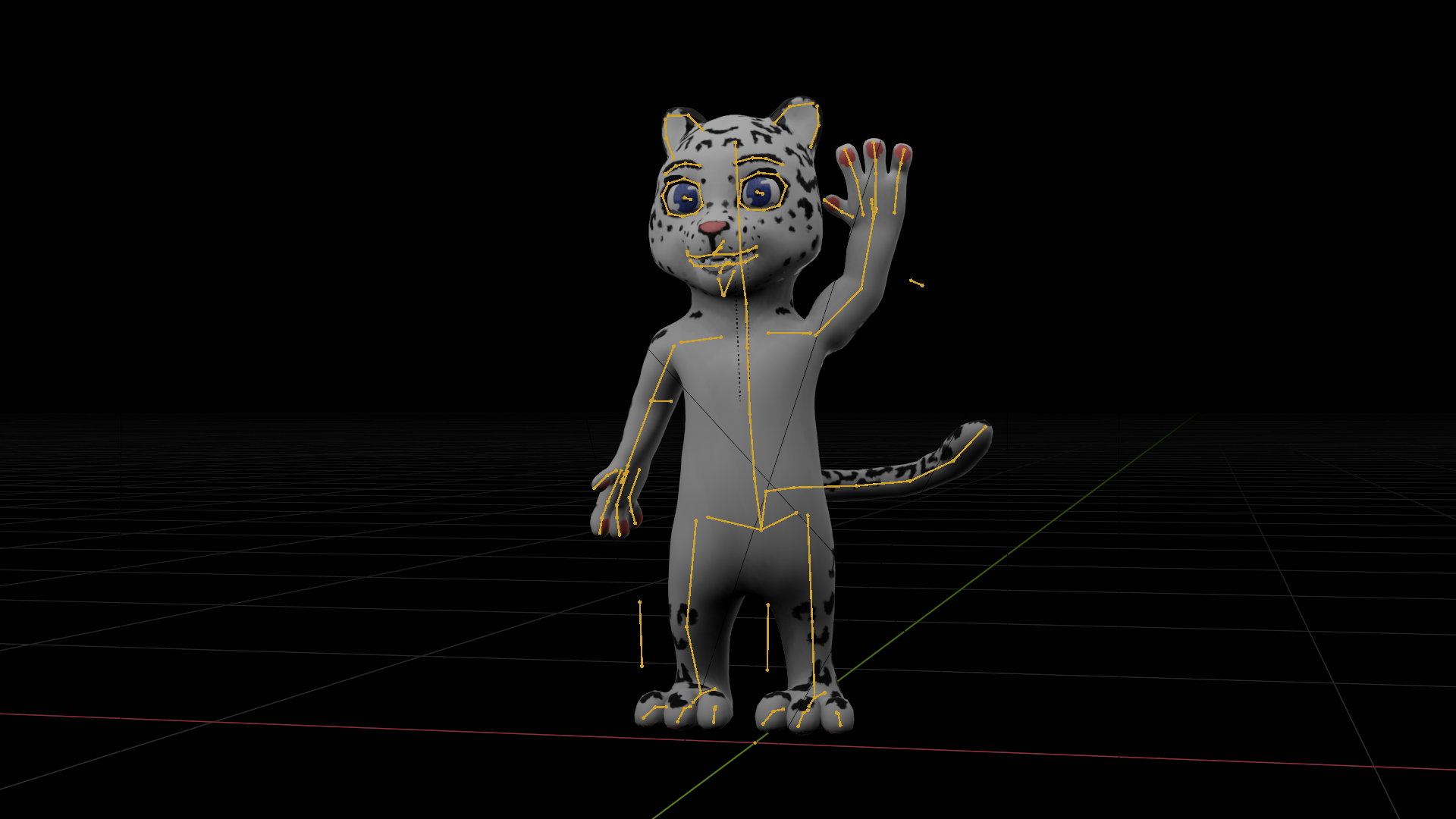}
\caption{Barsik says 'Salam!'}
\label{barsik}
\end{figure}

All answers are given by an animated 3D character, a new visualized assistant - \textbf{Barsik}. The model of Barsik \ref{barsik} and animations were developed at the Computer Science Department of Ala-Too International University. Barsik has her own unique design and represents the mascot of AIU. Additionally, we created a holography-style model of the university campus. To see the campus and information about buildings, HIVA user should give the voice command "studencheskiy gorodok" (transliterated translation of word campus).

Additionally, numeric and short responses to commands like "weather" or "news" is provided by displaying in the form of texts on a screen. It was designed for a better perception of the answer.

\begin{figure}
\centering
\includegraphics[width=0.45\textwidth]{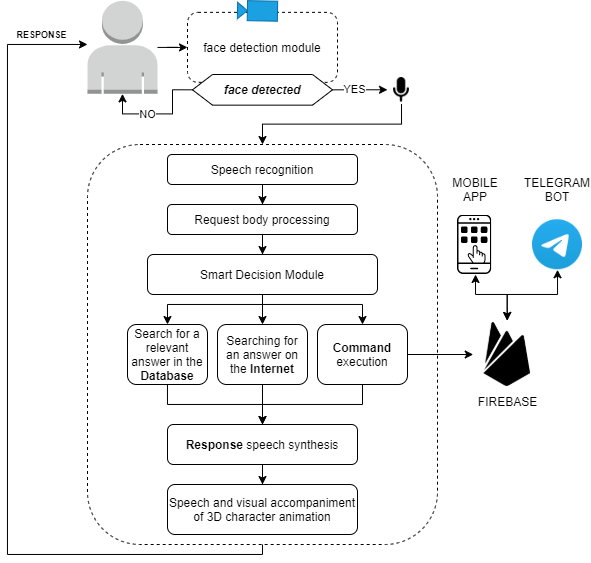} 
\caption{HIVA workflow design}
\label{fig4}
\end{figure}

\subsection{Software design and architecture}

\begin{figure}
\centering
\includegraphics[width=0.48\textwidth]{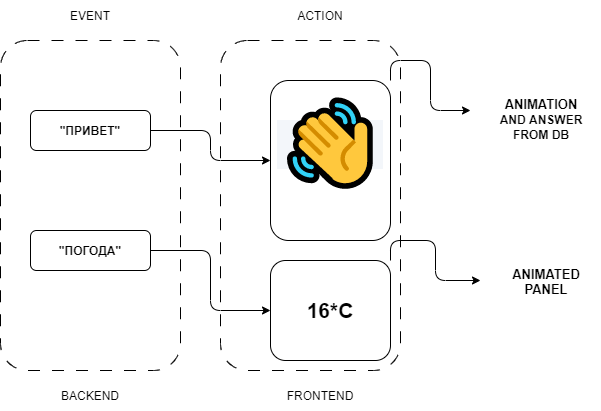}
\caption{Inter service interaction architecture}
\label{architecture}
\end{figure}

We needed to pick specific architecture to combine voice assistant and visual avatar in one system. As we all may know, there are many proposed designs and architectural patterns (cite patterns here). To connect two or more services as applications, we should refer to either a microservice pattern or an event-driven one \cite{Michelson}. We can assume users' voice commands as an event to pass them to be processed by the front-end part, where events can be handled and execute 3D avatar animations and display any information on boards and panels. The described process is shown schematically on \ref{architecture}, where you can see examples of two different events.

\subsection{Data mining and data augmentation}

Every voice assistant design faces the problem of forming a huge but efficient dataset for question answering. Suggestion categorization subsystem required at least 1000 supervised entries. Manual data collection, which included only voice suggestions from university students and professors, gave more than 200 entries. Voice data was batched through STT (Speech-to-text) tool. For those purposes, we used the Silero chatbot. Texts were manually corrected and labeled to form the base dataset. To enlarge that number of data samples, a data augmentation script was composed by us to suggest semantically similar terms by provided categories. Those terms were used to form the most frequent 3-grams. 

\subsection{Suggestions classification}

In HIVA text preprocessing we have used 1000 sentences (including augmented data), for feature extraction (building a simple BoW). In training a model, we have used the Multinomial Naive Bayes (MNB) algorithm, due to its simplicity in contrast to other sophisticated ML/NN models. There are three most popular Naive Bayes algorithms: Gaussian (for prediction of continuous variables), Bernoulli (for prediction of boolean values), Multinomial (for classification of categorical variables). 

As our aim was to classify requests into different categories, the MNB was the best matching algorithm among those three. 


We can see, that both metrics provide similar outputs: about 69.5\%. Thus, we assume that the initial model provides at least stable results on the data of our BoW. Further work will include expansion of the dataset, and implementation of other complex ML/NN algorithms and models.

\subsection{Short answers subsystem}

Another functional part of the HIVA application we designed was the short answers subsystem, which aimed to give short answers to user requests (for instance, question: how many buttons does the piano have? answer: 88). This interesting application feature was suggested to our students as Hackathon task. We provided a sample code with various test cases. Test cases included questions, URL pages to find answers in, and short answers. Sample code was able to pass 50\% of tests. The group of students managed to increase that result to 80\%. As we all know, modern Search engines, such as Google, Bing, and Yandex, can already provide answers by showing paragraphs  with main keywords highlighted before listing of websites. Yet another problem was to guess which URL would probably have an answer to the user's question. By now, we decided to use "otvet@mail.ru", a Russian analog of Quora question-answering service, as a data mining source. Since all information is open to everyone, we could run the script for extracting questions and applying a short answer extraction algorithm to the best answer text.

\begin{figure}
    \centering
    \includegraphics[width=0.48\textwidth]{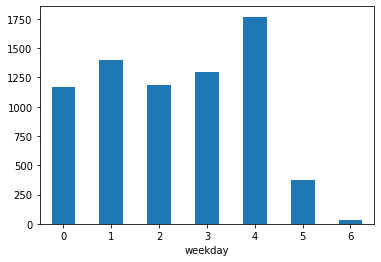}
    \caption{Amount of user requests per weekday}
    \label{fig:weekdayrequests}
\end{figure}

Version two of the HIVA project has been working continuously for one year, starting from September 2021 till now. Within this period, 7230 requests were made, with a mean daily usage value of 56.92, including weekends. You can observe total requests by weekdays on bar graph \ref{fig:weekdayrequests}, where 0 by the x-axis represents Monday. According to request body analysis, the most popular requests by now are "hello", "music" and "how are you?". Most frequent requests related to the university  about medical and humanities faculties, vice-rector, and tuition fees.


\section{Results and discussion}



This is one of the first attempts at NLP application in Kyrgyzstan. By now, HIVA NLP model performs relatively high accuracy in a wide range of 74-97\% on various tests. Therefore the efficiency of the model still has to be improved.




There are several potential points of growth: 
\begin{itemize}
\item Increase of the number of elements (rows) in the input dataset;
\item Reduction of features (columns) of the input dataset;
\item Work with different noises and accent recognition in a speech2text system;
\item Other code-related issues might be optimized
\end{itemize}

\section{Further work}


The HIVA as a typical virtual 3D assistant will provide the possibility for the expansion of the edge of the future for humanity. Combining such technologies as data science, machine learning, and 3D holography brings us to the next level of  person-to-machine interaction. Perhaps, virtual assistants such as Alexa, Siri, Cortana, and others - will receive their own 3D representations in the future. 

Also, there are other ways for HIVA extension, for example, the 6-DoF technologies: "Holographic video provides users with an immersive six degrees of freedom (6-DoF) viewing experience rather than traditional virtual reality (VR), 360 degrees, and other 3-DoF videos" \cite{Liu}. These technologies increase the immersive experience because 6-DoF - "allow users to walk around an object in a circle and view it from the top and the bottom" \cite{Liu}.

In the case of virtual assistants for HEIs, there might be other specific student-life-related needs: for example, issuing certificates on request. Also, the embedded  model of requests' classification in HIVA can be modified by increasing its accuracy. Today, the project team works on the model improvement and supports the HIVA. Finally, they are open to suggestions, partnerships, and collaborations with other HEIs, in work over HIVA and other projects, locally and abroad. 


\section*{ACKNOWLEDGMENTS}

We would like to express our sincere gratitude to Ala-Too International University for providing the grant that supported this research. Without their financial assistance, this study would not have been possible. Additionally, we would like to acknowledge the invaluable support and resources provided by the university, including access to laboratory facilities and equipment, and the expertise of their faculty members.



\begin{thebibliography}{00}

\bibitem{Abrahamsson} Abrahamsson, P., Salo, O.,
Ronkainen, J. \& Warsta, J. (2002) "Agile software development methods: Review and analysis," VTT
publication 478, Espoo, Finland, pp.63-91

\bibitem{Aggarwal} Aggarwal, C.C., Zhai, C. (2012). An Introduction to Text Mining. In: Aggarwal, C., Zhai, C. (eds) Mining Text Data. Springer, Boston, MA.

\bibitem{Caggianese} Giuseppe Caggianese, Giuseppe De Pietro, Massimo Esposito, Luigi Gallo, Aniello Minutolo, Pietro Neroni, "Discovering Leonardo with artificial intelligence and holograms: A user study," Pattern Recognition Letters, vol. 131, 2020, pp.361-367.

\bibitem{Cho} Cho, Juyun Joey, "An Exploratory Study on Issues and Challenges of Agile Software Development with Scrum," in All Graduate Theses and Dissertations. pp.12-36, 2010

\bibitem{Conner} Thomas Conner, "Pepper's Ghost and the augmented reality of modernity," Journal of Science \& Popular Culture, vol. 3, issue 1, pp. 57-79, Mar. 2020.

\bibitem{Duan} Duan, Hubert Haoyang, Vladimir G. Pestov, and Varun Singla. "Text categorization via similarity search." Proceedings of the 6th International Conference on Similarity Search and Applications-Volume 8199. 2013.

\bibitem{Girish} Girish, Patil, Shri Chhatrapati Shivaji, Pathade Omkar, and Dubey Shweta. "Holographic Artificial Intelligence Assistance." International Journal of Computer Applications 975, 2019, 8887.

\bibitem{Huang} Huang, Xinyu, Fridolin Wild, and Denise Whitelock. "Design dimensions for holographic intelligent agents: A comparative analysis." 2021, pp.1-10.

\bibitem{Joachims} Joachims, Thorsten. "Text categorization with support vector machines: Learning with many relevant features." Machine Learning: ECML-98: 10th European Conference on Machine Learning Chemnitz, Germany, April 21-23, 1998 Proceedings. Berlin, Heidelberg: Springer Berlin Heidelberg, 2005.

\bibitem{Liu} Z. Liu et al., "Point Cloud Video Streaming: Challenges and Solutions," IEEE Network, vol. 35, no. 5, pp. 202-09, Sept./Oct. 2021.

\bibitem{Mazyad} Mazyad, Ahmad, Fabien Teytaud, and Cyril Fonlupt. "A comparative study on term weighting schemes for text classification." Machine Learning, Optimization, and Big Data: Third International Conference, MOD 2017, Volterra, Italy, September 14-17, 2017, Revised Selected Papers 3. Springer International Publishing, 2018.

\bibitem{Patel} D. Patel and P. Bhalodiya, "3D Holographic and Interactive Artificial Intelligence System," 2019 International Conference on Smart Systems and Inventive Technology (ICSSIT), Tirunelveli, India, 2019, pp. 657-662.

\bibitem{Srivastava} A. Srivastava, S. Bhardwaj and S. Saraswat, "SCRUM model for agile methodology," 2017 International Conference on Computing, Communication and Automation (ICCCA), Greater Noida, India, 2017, pp. 864-869.

\bibitem{Sugimoto} Sugimoto, Maki. "Extended reality (XR: VR/AR/MR), 3D printing, holography, AI, radiomics, and online VR Tele-medicine for precision surgery." Surgery and Operating Room Innovation. Springer, Singapore, 2021. 65-70.

\bibitem{Wan} Y. -T. Wan, C. -C. Chiu, K. -W. Liang and P. -C. Chang, "Midoriko Chatbot: LSTM-Based Emotional 3D Avatar," 2019 IEEE 8th Global Conference on Consumer Electronics (GCCE), Osaka, Japan, 2019, pp. 937-940, doi: 10.1109/GCCE46687.2019.9015303.

\bibitem{Eiter} Eiter, Thomas, et al. "A Neuro-Symbolic ASP Pipeline for Visual Question Answering." Theory and Practice of Logic Programming 22.5, 2022, pp. 739-754.

\bibitem{Piscitello} Piscitello, Andrea, and Ettore Trainiti. "An Interactive method to control Computer Animation in an intuitive way," unpublished

\bibitem{Michelson} Michelson, B. M. (2006). "Event-driven architecture overview. Patricia Seybold Group," 2(12), 10-1571.

\bibitem{Sharma} Sharma, Priya, Sparsh Sharma, and Pooja Gambhir. "Prashn: University Voice Assistant." Artificial Intelligence and Speech Technology: Third International Conference, AIST 2021, Delhi, India, November 12-13, 2021, Revised Selected Papers. Cham: Springer International Publishing, 2022. 

\bibitem{Lemercier} J. Lemercier "Constellations - Studio Joanie Lemercier" Studio Joanie Lemercier - Light as a medium, space as a canvas. https://joanielemercier.com/constellations/ (accessed Feb. 28, 2023).

\bibitem{lucasquincoses} L. Quincoses Toledo "Lucas Quincoses Toledo (@lucasquincoses) - Sketchfab" Newsfeed - Sketchfab https://sketchfab.com/lucasquincoses (accessed Feb. 28, 2023).

\end{thebibliography}
\end{document}